\documentclass[epj]{webofc}
\usepackage[varg]{txfonts}   % Web of Conferences font
\usepackage{graphicx,color} % for figures
\usepackage{bm}       % bold math 
\usepackage{amsmath}  
\usepackage{amssymb}
 \woctitle{21st International Conference on Few-Body Problems in Physics}
\begin{document}
%%%%%%%%%%%%%%%%%%%%%%%%%%%%%%%%%%%%%%%%%%%%%
\renewcommand{\vec}{\boldsymbol}
\newcommand{\mc}{M_{\mathrm{c}}}
\newcommand{\mn}{M_{\mathrm{n}}}
\newcommand{\mnc}{M_{\mathrm{nc}}} 
\newcommand{\mr}{M_{\mathrm{R}}}
\newcommand{\gma}{\gamma}
\newcommand{\gmat}{\tilde{\gamma}}
\newcommand{\pc}{\vec{p}_{c}}
\newcommand{\pn}{\vec{p}_{n}}
\newcommand{\bes}{{}^{7}\mathrm{Be}}
\newcommand{\besst}{{}^{7}\mathrm{Be}^{\ast}}
\newcommand{\be}{{}^{8}\mathrm{B}}
\renewcommand{\S}[2]{{}^{#1}S_{#2}}
\renewcommand{\P}[2]{{}^{#1}P_{#2}}
\newcommand{\gone}{g_{(\S{3}{1})}}
\newcommand{\gtwo}{g_{(\S{5}{2})}}
\newcommand{\gthree}{g_{(\S{3}{1}^{*})}}
\newcommand{\aone}{a_{(\S{3}{1})}}
\newcommand{\atwo}{\ensuremath{a_{(\S{5}{2})}}}
\newcommand{\hone}{h_{(\P{3}{2})}}
\newcommand{\htwo}{h_{(\P{5}{2})}}
\newcommand{\hpt}{h_{t}}
\newcommand{\hthree}{h_{(\P{3}{2}^{*})}}
\newcommand{\honet}{\tilde{h}_{(\P{3}{1})}}
\newcommand{\htwot}{\tilde{h}_{(\P{5}{1})}}
\newcommand{\Xone}{{X}_{(\S{3}{1})}}
\newcommand{\Xtwo}{{X}_{(\S{5}{2})}}
\newcommand{\Xonet}{\tilde{X}_{(\S{3}{1})}}
\newcommand{\Xtwot}{\tilde{X}_{(\S{5}{2})}}
\newcommand{\V}[1]{\vec{V}_{#1}}
\newcommand{\fdu}[2]{{#1}^{\dagger #2}}
\newcommand{\fdd}[2]{{#1}^{\dagger}_{#2}}
\newcommand{\fu}[2]{{#1}^{#2}}
\newcommand{\fd}[2]{{#1}_{#2}}
\newcommand{\T}[2]{T_{#1}^{\, #2}}
\newcommand{\e}{\vec{\epsilon}}
\newcommand{\es}{\e^{*}}
\newcommand{\cw}[2]{\chi^{(#2)}_{#1}}
\newcommand{\cwc}[2]{\chi^{(#2)*}_{#1}}
\newcommand{\cwf}[1]{F_{#1}}
\newcommand{\cwg}[1]{G_{#1}}

\newcommand{\ke}{k_{E}}

\newcommand{\kest}{k_{E\ast}}

\newcommand{\kc}{k_{C}}
\newcommand{\upartial}[1]{\partial^{#1}}
\newcommand{\dpartial}[1]{\partial_{#1}}
\newcommand{\etae}{\eta_{E}}
\newcommand{\etab}{\eta_{B}}
\newcommand{\etaest}{\eta_{E\ast}}
\newcommand{\etabst}{\eta_{B\ast}}
\newcommand{\vecpt}[1]{\hat{\vec{#1}}}
\newcommand{\uY}[2]{Y_{#1}^{#2}}
\newcommand{\dY}[2]{Y_{#1 #2}}

\def\lsim{\mathrel{\rlap{\lower4pt\hbox{\hskip1pt$\sim$}}
    \raise1pt\hbox{$<$}}}         %less than or approx. symbol
\def\gsim{\mathrel {\rlap{\lower4pt\hbox{\hskip1pt$\sim$}}
    \raise1pt\hbox{$>$}}}         %greater than or approx. symbol
%%%%%%%%%%%%%%%%%%%%%%%%%%%%%%%%%%%%%%%%%%%%%%%%%%%%%%%%%%%%%%%%%%%%%%%
%
\title{How well do we understand ${}^7{\rm Be} + p \rightarrow {}^8{\rm B} + \gamma$? \\ An Effective Field Theory perspective 
%\footnote{Contribution to the $21^\text{st}$ International Conference on Few-Body Problems in Physics}
}

\author{Xilin Zhang \inst{1,2}\fnsep\thanks{\email{xilinz@uw.edu}} \and
        Kenneth M.~Nollett\inst{3,4,2}\fnsep\thanks{\email{nollett@mailbox.sc.edu}} \and
        D.~R.~Phillips \inst{2}\fnsep\thanks{\email{phillid1@ohio.edu}}
        % etc.
}

\institute{Physics Department, University of Washington, 
Seattle, WA 98195, USA
\and
         Institute of Nuclear and Particle Physics and Department of
Physics and Astronomy, Ohio University, Athens, OH\ \ 45701, USA
\and
         Department of Physics, San Diego State University,
5500 Campanile Drive, San Diego, California 92182-1233, USA
\and 
      Department of Physics and Astronomy, 
University of South Carolina,
712 Main Street, Columbia, South Carolina 29208, USA  }

%
%\author{Xilin Zhang} \email{xilinz@uw.edu}
%\affiliation{Physics Department, University of Washington, 
%Seattle, WA 98195, USA} 
%\affiliation{Institute of Nuclear and Particle Physics and Department of
%Physics and Astronomy, Ohio University, Athens, OH\ \ 45701, USA}
%
%\author{Kenneth M.~Nollett} \email{nollett@mailbox.sc.edu}
%\affiliation{Department of Physics and Astronomy, 
%University of South Carolina,
%712 Main Street, Columbia, South Carolina 29208, USA} 
%\affiliation{Department of Physics, San Diego State University,
%5500 Campanile Drive, San Diego, California 92182-1233, USA} 
%\affiliation{Institute of Nuclear and Particle Physics and Department of
%Physics and Astronomy, Ohio University, Athens, OH\ \ 45701, USA}
%
%
%\author{D.~R.~Phillips} \email{phillid1@ohio.edu}
%
%\affiliation{Institute of Nuclear and Particle Physics and Department of
%Physics and Astronomy, Ohio University, Athens, OH\ \ 45701, USA}

\abstract{
We have studied the $^7\mathrm{Be}(p,\gamma)^8\mathrm{B}$ reaction in the Halo effective field theory (EFT) framework. The leading order (LO) results were published in Ref.~\cite{Zhang:2014zsa} after the isospin mirror process, $^7\mathrm{Li}(n,\gamma)^8\mathrm{Li}$, was addressed in Ref.~\cite{Zhang:2013kja}. In both calculations, one key step was using the final shallow bound state asymptotic normalization coefficients (ANCs) computed by {\it ab initio} methods to fix the EFT couplings. Recently we have developed the next-to-LO (NLO) formalism~\cite{Zhang:2015}, which could reproduce other model results by no worse than $1\%$ when the $^7\mathrm{Be}$-p energy was between $0$ and $0.5$ MeV. In our recent report~\cite{Zhang:2015ajn}, a different approach from that in Ref.~\cite{Zhang:2014zsa} was used. We applied Bayesian analysis to constrain all the NLO-EFT parameters based on measured $S$-factors, and found tight constraints on the $S$-factor at solar energies. Our $S\left(E=0\,\mathrm{MeV}\right)= 21.3\pm 0.7$ \mbox{eV b}. The uncertainty is half of that previously recommended. In this proceeding, we provide extra details of the Bayesian analysis, including the computed EFT parameters' probability distribution functions (PDFs) and how the choice of input data impacts final results. 
}

%\begin{abstract}
%\end{abstract}
%
\maketitle
\section{Introduction} \label{intro}
The $^7\mathrm{Be}(p,\gamma)^8\mathrm{B}$ reaction cross section around zero energy directly affects the solar neutrino flux \cite{Robertson:2012ib}. 
(An expanded discussion on the reaction's importance for astrophysics can be found in Ref~\cite{Zhang:2015ajn}.)
The previously recommended $S(0) = 20.8 \pm 0.7\pm 1.4$ $\mbox{eV b}$ \cite{Adelberger:2010qa}. The first error comes from the capture experiment uncertainties. Due to the Coulomb repulsion barrier, the cross section is exponentially suppressed, so that the lowest energy of available data is above $0.1$ MeV. Extrapolating the data down to zero energy based on various models gives rise to the second quoted $S(0)$ error. Therefore eliminating the model dependence is critical for improvement. 

Fortunately different scales in this reaction are well separated, and thus the problem is ``simple'' enough to be treated in Halo EFT. The $^7\mathrm{Be}$ and proton are considered as two fundamental particles and $^8$B as a shallow bound state of the two; the high momentum modes resolving the short distance ($< 3$ fm) physics  are integrated out, and their effects are subsumed to contact terms in the lagrangian. Although the idea is close to the previous potential models', the EFT puts the scale separation on rigorous grounds and should increase in accuracy as terms of higher order in momentum are retained. The number of parameters increases while going to higher order, but it is kept minimal, and the error of truncation can be estimated. Moreover the important effect of the  Coulomb interaction between $^7$Be and proton can be treated exactly. The reaction is entirely an electric-dipole ($E1$) induced transition from the $s$- and $d$-wave initial states to  $p$-wave configurations in the final $^8$B.  
The resulting $S$-factor \cite{Zhang:2015} is proportional to 
\[\textstyle{ S(E)\propto \sum_{s} C_{s}^2 
\bigg[ \big\vert \mathcal{S}_\mathrm{EC} \left(E;a_s,r_s\right) 
+ \overline{L}_{s} \mathcal{S}_\mathrm{SD} \left(E;a_s,r_s\right)  
+ \epsilon_{s} \mathcal{S}_\mathrm{CX}\left(E;a_s,r_s\right) \big\vert^2 
+|\mathcal{D}_\mathrm{EC}(E)|^2 \bigg]} \ . 
\] 
The proportionality factor is in Ref.~\cite{Zhang:2014zsa}. The total spin $s$ of $^7\mathrm{Be}$-p can be $1$ or $2$, defining two different reaction channels. The LO term $\mathcal{S}_\mathrm{EC}$ is the amplitude of the external direct capture to the $^7\mathrm{Be}$-p component in $^8$B; the same capture but to the lowest excited core $^7\mathrm{Be}^\ast$ plus proton component is a NLO contribution denoted as $\epsilon_s \mathcal{S}_\mathrm{CX}$; thus the dimensionless factor $|\epsilon_s|$ is small. $\overline{L}_s \mathcal{S}_\mathrm{SD}$ is another NLO term coming from the capture in short distance ($< 3$ fm); the coefficient $\overline{L}_s \sim 3$ fm.  The three amplitudes in each channel depend on the s-wave scattering length $a_s$ and effective range $r_s$ due to the initial state interaction effect. $\mathcal{D}_\mathrm{EC}$ is the $E1$ transition from the d-wave initial state to the final bound state, which is not affected by the initial state interaction at LO. The overall factor in each channel, $C_s$, is the ANC of the two-particle wave function (with total spin $s$) in $^8$B.  

There are nine EFT parameters, $C_{1,2}^2$, $a_{1,2}$, $r_{1,2}$, $\overline{L}_{1,2}$, and $\epsilon_1$ ($\epsilon_2=0$ based on angular momentum conservation). Their so-called posterior PDF can be extracted from direct capture data by using Bayesian analysis \cite{SiviaBayesian96}. 
In our analysis, another five parameters $\xi_{1,2,3,4,5}$ were introduced to float the normalizations of the five capture data sets used. (In total 42 data points with energies up to $0.5$ MeV were used; Fig.~\ref{fig-2} provides some information about them.)  A Markov chain Monte Carlo (MCMC) algorithm with Metropolis sampling \cite{SiviaBayesian96} was used to compute the 14-dim posterior PDF~\cite{Zhang:2015ajn}. 
Note prior knowledge about the parameters was  incorporated by assigning a prior PDF to each of them. For $a_{1,2}$, it was a Gaussian distribution centered at the measured value~\cite{Angulo2003a} with the width equal to the experimental uncertainty; the other EFT parameters were given flat  prior PDFs over finite value windows \cite{Zhang:2015ajn}. The  $\xi_i$ priors were also Gaussian distributions but centered at $0$ (the data set's normalization factor is $1+\xi_i$); the widths were the corresponding experimental common mode errors (CMEs), $2.7\%$, $2.3\%$, $11.25\%$, $5\%$, and $2.2\%$ in the order of being mentioned in Fig.~\ref{fig-2}. (Ref.~\cite{Hammache:2001tg} contains mostly relative cross sections, so its CME is handled differently.)

\section{EFT parameters}
\label{sec-1}

\begin{figure}
% Use the relevant command for your figure-insertion program
% to insert the figure file.
\centering
\includegraphics[width=14cm,clip]{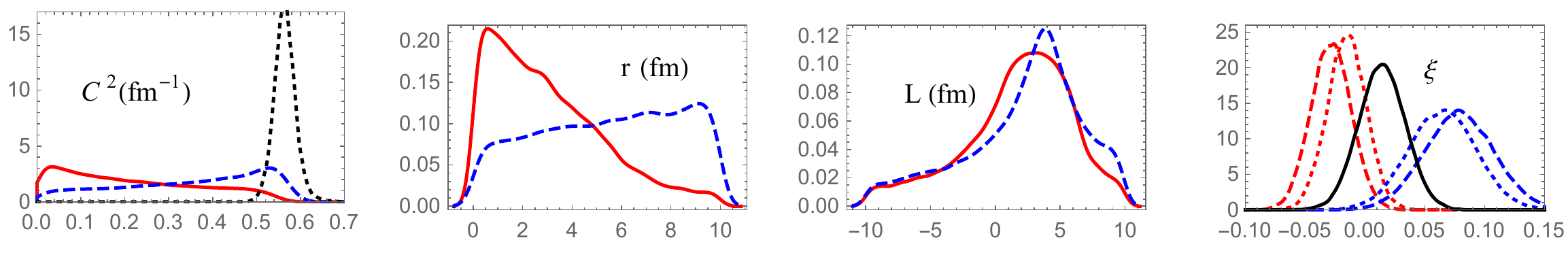}
\caption{In the first panel, the red solid, blue dashed, and black dashed lines are the posterior PDFs of $C_1^2$, $C_2^2$, and $C_1^2 + C_2^2$; in the second (third) panel, the red solid and blue dashed lines are the PDFs of $r_s$ ($\overline{L}_s$) in $s=1$ and $2$ channels, respectively; in the last panel, the red dashed (for the Junghans BE1 data set), red dotted (the Junghans BE3), blue dashed (Filippone), blue dotted (Hammache 1998), and black solid (Baby) lines are the PDFs of the corresponding data normalization variable $\xi_i$. The data sets are shown in Fig.~\ref{fig-2}.}
\label{fig-1}       % Give a unique label
\end{figure}

%A few key input parameters . The mass of $\bes$ is $6534.18$ MeV, and the $\be$ proton separation energy $0.1364$ MeV \cite{Wang2012}. 
Fig.~\ref{fig-1} shows a few interesting 1-dim parameter posterior PDFs. (Each PDF is the 14-dim PDF with the other 13 parameters marginalized, and was computed via drawing its histogram based on the MCMC samples~\cite{Zhang:2015ajn}.) We see in the first panel that the data prefer a large $C_2^2$ and a small $C_1^2$,  and the sum is tightly constrained ($C_1^2 +C_2^2 = 0.564(23)\, \mathrm{fm}^{-1}$~\cite{Zhang:2015ajn}). The second panel shows the $^7$Be-p s-wave effective ranges. Although we assigned positive flat priors for them, we also required that no s-wave resonance exists below $0.6$ MeV \cite{Zhang:2015ajn}. The computed PDF of $r_1$  disfavors large $r_1$ value, while that of  $r_2$'s is almost flat. (The scattering lengths' PDFs are close to their priors.) In the third panel, each $\overline{L}_s$ distribution is broad but does have a significant peak around a few fm, overlapping with the $\overline{L}_{1,2} \sim 3$ fm region suggested by the EFT power counting. Thus we believe that we have detected the direct capture at short distance from the data, although the large distance capture dominates the reaction. The last panel shows the PDFs of $\xi_i$. The corresponding data sets are plotted in Fig.~\ref{fig-2}, which shows minor tension between the recent Junghans data sets and the rest. This is also reflected in the $\xi_i$ PDFs in the last panel: reducing the Junghans BE1 and BE3 normalizations is favored, the Filippone and Hammache normalizations need to be increased a bit, and Baby's is not modified significantly. However all floating norms (i.e. $1+\xi_i$), while $\neq 1$, are entirely consistent with the quoted CMEs.

\section{ S factor}

\begin{figure}
\centering
\includegraphics[width=7cm,clip]{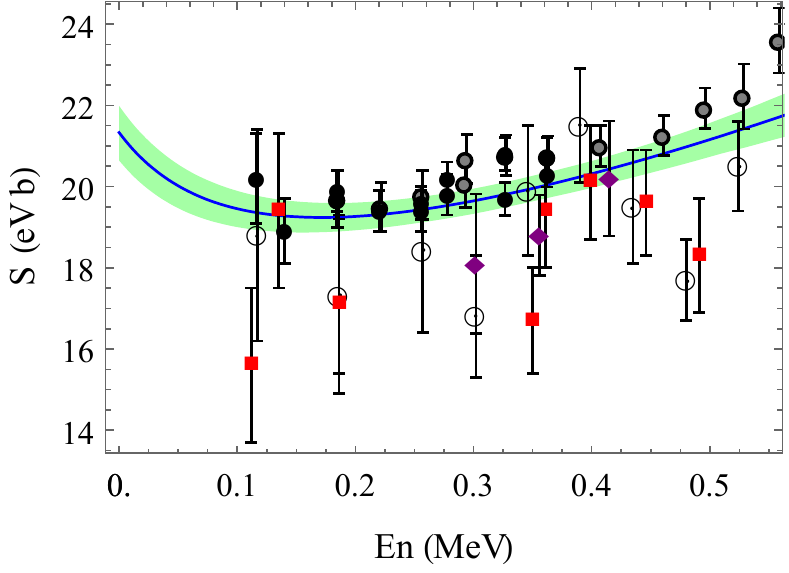}
\includegraphics[width=7cm,clip]{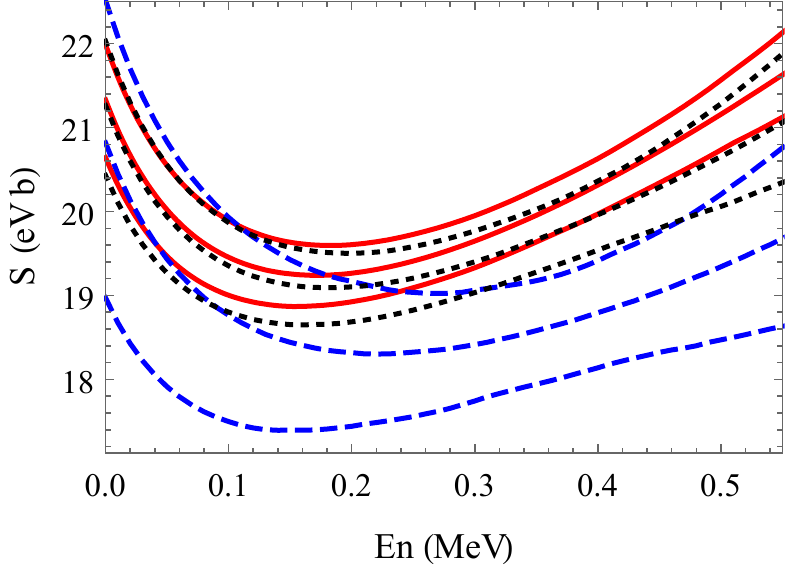}
\caption{The left panel shows the $S$-factor median values (solid blue
  curve) and the $1$-$\sigma$ intervals (shaded in green) at different energies.  
	The data used in our analysis and a few others above $0.5$ MeV are shown, including 
 Junghans {\it et.al.}, BE1 and BE3 \cite{Junghans:2010zz} (filled
black circle and filled grey circle), Filippone {\it et.al.,}
\cite{Filippone:1984us} (open circle), Baby {\it et.al.,}
\cite{Baby:2002hj, Baby:2002ju} (filled purple diamond), and Hammache
     {\it et.al.,} \cite{Hammache:1997rz, Hammache:2001tg} (filled red box).	
In the right panel, we show three different sets of the $S(E)$ median values and  $1$-$\sigma$ bounds, as computed by using different combinations of the data sets. The blue dashed lines use the combination of Filippone, Hammache and Baby sets, the black dashed lines add the Junghans BE3 set, and the red solid lines use all the sets.}
\label{fig-2}       % Give a unique label
\end{figure}

The left panel in Fig.~\ref{fig-2} shows the computed $1$-$\sigma$ lower and upper bounds and the median value for the $S$ factor. The three quantities were inferred from the $S$-factor distributions at individual energies, which were computed via making the histograms of $S$ based on the MCMC samples. The data are plotted for comparison. At the shown energies, the $S$ factor uncertainties $\approx 3.5\%$ or less, and are correlated. A numerical table can be found in Ref.~\cite{Zhang:2015ajn}. It is interesting to ask how the choice of data sets impacts our results; this is illustrated in the right panel. Without having the BE1 and BE3 sets in the  analysis, the $S$ factor uncertainty $\approx 9\%$ at zero energy; adding either BE1 or BE3 narrows the error bands significantly within the bounds provided by other data.
 Note that all the three $1$-$\sigma$ intervals overlap with each other.  We also see that combining BE1 with the other four sets does not change the constraints below $0.1$ MeV much; at higher energy the BE1's impact shows up by constraining $S(E)$ further and systematically favoring larger $S(E)$. 

We also repeated our analysis using different priors for the scattering lengths and introducing a N$^2$LO term into the capture amplitude, and we found negligible differences. We also found that the current $0.7\%$ uncertainty of $^8\mathrm{B}$'s proton separation energy could change $S(0)$ by about $0.75\%$.

\section{Summary}
In summary, the $^7\mathrm{Be}(p,\gamma)^8\mathrm{B}$ reaction has been studied in Halo EFT  up to NLO. We applied Bayesian analysis to constrain the EFT parameters. The analysis gave  a stringent constraint on the total ANC squared, and showed that positive short-distance-contribution parameters $\overline{L}_{1,2}$ were preferred. Minor tensions among the modern capture data sets were seen. We found that $S(E)$ is constrained to $\approx	 3.5\%$ between 0 and 0.5 MeV. The $S(0)$ results were robust against exclusion of either Junghans data set and against addition of an N$^2$LO term to the reaction amplitude.

\begin{acknowledgement}
X.Z.~and D.R.P.~acknowledge support from the US Department of Energy under grant DE-FG02-93ER-40756. X.Z. also acknowledges support from the US Department of Energy under grant DE-FG02-97ER-41014. K.M.N.~acknowledges support from the Institute of Nuclear and Particle Physics at Ohio University, and from U.S.~Department of Energy Awards No.~DE-SC 0010 300 and No.~DE-FG02-09ER41621 at the University of South Carolina.
\end{acknowledgement}

%\section*{Acknowledgements}

%
% BibTeX or Biber users please use (the style is already called in the class, ensure that the "woc.bst" style is in your local directory)
%\bibliographystyle{apsrev}

 \bibliography{./nuclear_reaction-7-26-2015}

\end{document}